\documentclass[11pt]{elsart}
\usepackage{setspace}
\usepackage{cases}
\usepackage[dvips]{graphicx}

\begin{document}
\begin{frontmatter}

\title{Effect of correlations on controllability transition in network control}
\author{Sen Nie$^a$}
\author{Xuwen Wang$^{a}$}
\footnote{Email:
xww86@mail.ustc.edu.cn}
\author{Binghong Wang$^{a,b,c}$}
 \footnote{Email: bhwang@ustc.edu.cn}
\address{$^{a}$ Department of Modern Physics,
University of Science and Technology of China,  Hefei, 230026, P. R.
China}
\address{$^{b}$ College of Physics and Electronic Information Engineering, Wenzhou
University, Wenzhou£¬ Zhejiang  325035, P. R. China}
\address{$^{c}$ School of Science, Southwest University of Science and Technology,
Mianyang, Sichuan, 621010, P. R. China}

\begin{abstract}
The numerical controllability transition makes the success of control can be achieved by increasing the number of driver nodes to a certain point. Motivated by the fact that the degree correlation has vast role in the dynamics on networks,
we study the impact of various degree correlations of different networks on the controllability transition point and find that the transition point depicts local maximum in sparse networks as degree correlation $r$ around 0.1 and 0 in ER and SF networks respectively. With the increasing of average degree, the local maximum disappear and the controllability transition cannot be influenced by degree correlation and degree distribution in dense ER networks. The results are supported by numerical simulations and provide more details to estimate the minimal driver nodes in large networks.

%
%
%
\end{abstract}

\date{}
\end{frontmatter}


\section{Introduction} \label{sec:intro}
The theory of control has been combined with complex networks recently and it attempts to solve the ultimate problem what is how to guide the system's behavior towards a desired state with external inputs in finite time\cite{liu2011controllability,liu2013observability,nepusz2012controlling,yuan2013exact}. Similar to the study in classical control theory, we firstly should judge whether a system is controllable or not. However the difficulty is rooted in the fact that it is hard to judge and compute the controllability of huge nonlinear system. Based on this, Liu et al. simplified the problem and studied the structural controllability of system to bypass the need of link weights in calculation of Kalman's controllability rank condition\cite{liu2011controllability}, and they proposed a method to identify the minimum driver nodes to steer the whole system. Then Ref\cite{yuan2013exact} studied the exact controllability of complex network with consideration of link weights, and many researchers began to focus on the analysis, optimal and application of controllability of complex networks\cite{posfai2013effect,cowan2012nodal,wang2012optimizing,liu2012control,yan2012controlling,ruths2014control,jia2013control,zhao2014universal,li2014controllability,jia2014connecting,pan2014structural,nie2014robustness}.

However, Ref\cite{sun2013controllability}found the large system is still uncontrollable when it meets the Kalman's controllability rank condition already, that is because the controllability Gramian is still ill conditioned. And it pointed out that the control success rate is enhanced and has a sharp transition by increasing the number of control inputs. After that, some researches on the controllability of complex networks based on the Gramian have appeared\cite{cornelius2013realistic,enyioha2014controllability,pasqualetti2014controllability,summers2013optimal}, and they discussed the bounds of energy and eigenvalue, limitation of computation and depict the Gramian matrix by some indexes. Unfortunately, there still lacks of some criteria to compute the minimum control inputs based on the controllability Gramian.

In this paper, we pay attention to the controllability transition point and study the relationship between the system's architecture and its controllability under minimum energy control trajectories. The results shows the transition point presents local maximum around degree correlation $r=0.1$ and $r=0$ in sparse ER and SF networks respectively; and it depends slightly on the degree correlation in dense networks.
The rest of this paper is organized as follows. Section II is the model and problem formulation, and we present our main results and analysis in Section III, followed by conclusion in Section IV.

\section*{Model and Problem Statement}
Consider a N-dimensional linear time-invariant dynamical system:

\begin{equation}\label{eq1}
\centering
  \dot{x}(t)=Ax(t)+Bu(t),
\end{equation}
where $x(t)=(x_{1}(t),x_{2}(t),\cdot\cdot\cdot,x_{N}(t))^{T}$ is the state of system at time $t$, $A$ is adjacent matrix which represents the interaction strength between nodes, $B$ is input matrix which defines how the input signals are connected to the nodes of networks and $u(t)=(u_{1}(t),u_{2}(t),\cdot\cdot\cdot,u_{M}(t))^{T}$ is the input vector. Corresponding to the minimal energy control trajectories in $t\in [t_{0},t_{1}]$:

\begin{equation}\label{eq2}
\centering
  u(t)=B^{T}\Phi^{T}(t_{0},t)W^{-1}(t_{0},t_{1})[\Phi(t_{0},t_{1})x^{(1)}-x^{(0)}],
\end{equation}

\begin{equation}\label{eq3}
\centering
  x(t)=\Phi(t,t_{0})[x^{(0)}+M_{t_{0},t_{1},t}(\Phi(t_{0},t_{1})x^{(1)}-x^{(0)})],
\end{equation}
where $M_{t_{0},t_{1},t}=W(t_{0},t)W^{-1}(t_{0},t_{1})$,
$W(t_{0},t_{1})=\int_{t_{0}}^{t_{1}}\Phi(t_{0},t)BB^{T}\Phi^{T}(t_{0},t)\,dt$ is the controllability Gramian and $\Phi(t_{1},t_{0})=e^{(t_{1}-t_{0})A}$\cite{rugh1996linear}.
Usually, according to the Kalman's rank condition\cite{kalman1963mathematical}, if the matrix $K=[B,AB,A^{2}B,\cdots, A^{N-1}B]$ has full rank, then system $(A,B)$ is considered controllable, vice versa. But Ref\cite{sun2013controllability} points that a system has a well conditioned Kalman's controllability matrix may has an ill conditioned controllability Gramian in large networks, because the Gramian matrix could be nearly singular with less driver nodes. However, according to the Gramian matrix of the criteria of controllability, the system is uncontrollable when Gramian matrix is singular\cite{rugh1996linear}.
Therefore they characterize the difficulty of calculating the inverse of $W$ by the reciprocal condition number $\gamma(W)$, and give an example as a directed linear chain containing $N$ nodes without self-loops. According to Kalman's rank condition, the chain can be controlled by the first node, however, the $\gamma(W)$ decreases exponentially as a function of the number of nodes $N$ , which leads to the numerical control of system impossible. As the driver nodes increased,
there is a controllability transition of numerical success rate increases from zero to one, and the driver nodes corresponding to the transition of success rate is called controllability transition point.

It is already known that the degree distribution and the degree correlation affect the structural controllability of networks\cite{liu2011controllability,posfai2013effect}, and Ref\cite{sun2013controllability}has demonstrated the impact from degree distribution to the controllability transition. Based on this foundation, we consider whether and how the degree correlation affects the controllability transition. Firstly, we motivate our work by comparing the effect of degree correlation to the transition point in different networks, then we give the further analysis and more details.

\section*{Methods and Results }
Each node in network has an in-degree $k_{in}$ and out-degree $k_{out}$. The degree correlations is the correlations between the source node¡¯s in- and out- degree, and the target node¡¯s in- and out- degree. It can be quantified by the Pearson coefficient\cite{foster2010edge}:

\begin{equation}\label{eq4}
\centering
  r=\frac{M^{-1}\sum_{i}j_{i}k_{i}-[M^{-1}\sum_{i}1/2(j_{i}+k_{i})]^{2}}{M^{-1}\sum_{i}1/2(j_{i}^{2}+k_{i}^{2})-[M^{-1}\sum_{i}1/2(j_{i}+k_{i})]^{2}},
\end{equation}
where $\sum_{i}\cdot$ sums over all edges, $j$ and $k$ are the in- or out- degrees of source and target nodes respectively which belong to edge $i$ for directed networks. For undirected networks, $j$ and $k$ are the degrees of two nodes belong to edge $i$. Positive value of $r$ indicates the assortative network while the negative value of $r$ characterize the disassortative network.

We start from the undirected Erd\H{o}s-R\'{e}nyi (ER) and Scale-free (SF) networks with self-loops. The edges and diagonal elements $A_{ii}$ are assigned weights drawn from an uniform distribution in $[-1,1]$. $B$ is diagonal matrix whose diagonal element $B_{ii}$ is assigned $1$ when the $i-$th node is driver node. The initial state and target state are given in our networks, and $f$ nodes are randomly chosen as driver nodes at each independent realization. We increase the driver nodes $f$ and consider the numerical control successful if the distance $\eta$ between calculated state $x(t_{1})$ and target state $x^{(1)}$ is less than $10^{-6}$ in time window $t_{0}\leq t \leq t_{1}$, where $\eta\ll\|x^{(1)}-x^{(0)}\|$. Then the driver nodes corresponds to the point which success control firstly happens is the transition point $n_{d}$.

 Simulated annealing is used to obtain the networks with desired degree correlation by rewiring links, while leaving the in- and out- degree unchanged. By tuning the degree correlation $r$ to different desired values, we compute the transition point in different networks and try to find out the relationship between the degree correlation and the controllability transition.

\begin{figure}
\centering{\includegraphics[height=2.8in,width=8.5in]{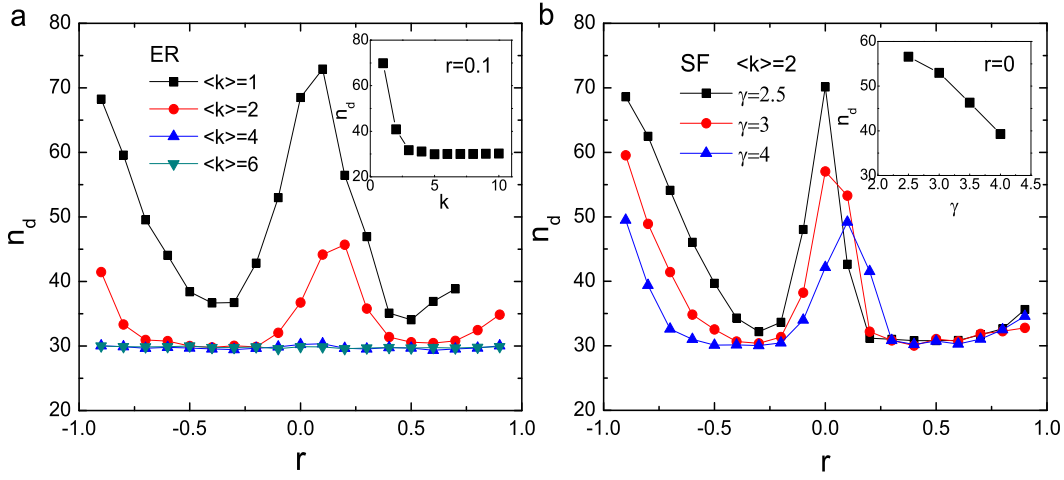}}
\caption{(Color online) The impact of degree correlations on the controllability transition points for ER and SF networks with $N=100$. In both cases the initial states are chosen randomly on the unit sphere centered at the origin and the target states are randomly oriented $\sigma=10^{-2}$. Insets: the  transition point as the function of the average degree as $r=0.1$ for ER networks[panel(a)]; the  transition point as the function of the degree exponent as $r=0$ for SF networks[panel(b))]. Each data point is an average of 100 independent realizations.
} \label{fig1}
\end{figure}

Firstly, we investigate how the degree correlation affect the controllability transition point. Fig.~\ref{fig1} features the minimum number of driver nodes that needed to achieve the numerical success control as a function of degree correlation.
Obviously, the curves of the transition point in ER networks could be divided into two categories: (i) Sparse networks: The value of transition points reach the local maximum as $r\in (0.1,0.2)$ and they are also large as $r$ around -1. In this sense, the most disassortative and less assortative ER networks are harder to control. But interestingly, the ER networks are easier to control as $r$ around 0.5 and -0.5.
(ii) Dense networks: The transition point slightly depends on the degree correlation, which is greatly unlike the situation in sparse networks. In other words, the average degree impact greatly on the transition point. However, the large average degree weakens the effect of degree correlation on the networks controllability, which can be seen in the inset of Fig.~\ref{fig1}(a).
 Meanwhile, there is similar situation for SF networks whose value of transition point reaches the local maximum as $r$ around $0$ and the transition point corresponding to the local maximum shifts right with the increasing of degree exponent $\gamma$. The inset of Fig.~\ref{fig1}(b) indicates the local maximal value of transition point decreasing with $\gamma$ increased.

According to the curves in Fig.~\ref{fig1}, we also find the minimal value of transition point could not be influenced greatly by degree distribution in dense networks. Surprisingly, it is different from the results in structural controllability which the number of driver nodes decreases drastically as the increasing of average degree and degree exponent. Since the network is considered successfully controlled by increasing driver nodes till the distance between target state and the calculated state at final time satisfies $\|x^{(1)}-x(t_{1})\|<\eta$, then we could get $n_{d}\geq n_{dc}$, where $n_{dc}$ is a critical value determined by $t_{1}-t_{0}$,$x^{(0)}$,$x^{(1)}$,$A$ and the control mode. The results 
shows the number of transition point increasing as network size linearly in our simulations.

\begin{figure}
\centering{\includegraphics[height=2.8in,width=8.5in]{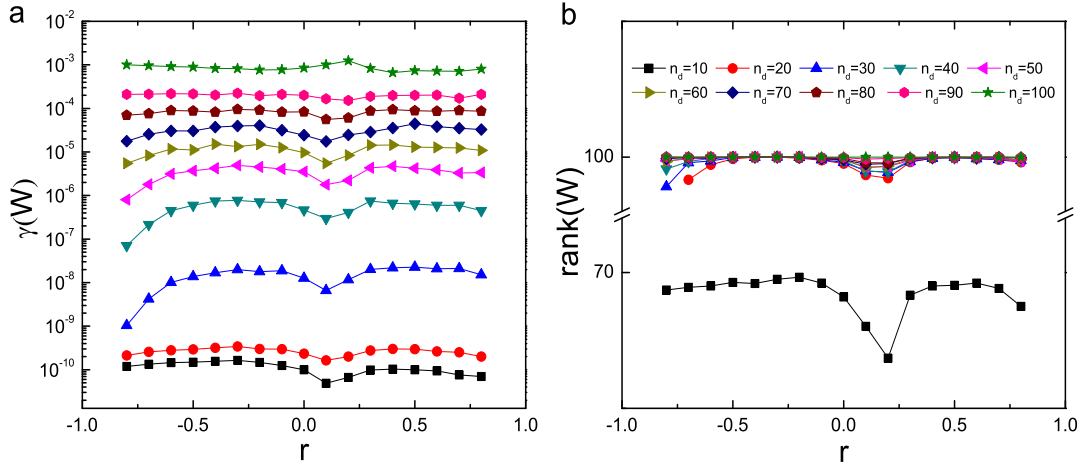}}
\caption{(Color online) Numerical $\gamma(W)$ and $rank(W)$ with the change of degree correlation under different driver nodes for undirected ER networks as $k=2$. The statistics and parameters not shown are the same used in Fig.1.
} \label{fig3}
\end{figure}

In order to explore the reason for the local maximal transition point in sparse networks, we examine the reciprocal condition number $\gamma(W)$ (the ratio of the smallest to the largest singular value) and $rank(W)$. The results shown in Fig.~\ref{fig3} depicts the change of $\gamma(W)$ with different degree correlations under various number of driver nodes for undirected ER  networks as $<k>=2$. Fig.~\ref{fig3} intuitively shows there are reduction area in curves around $r=0.1$ both in $\gamma(W)$ and $rank(W)$, where the local maximal transition point appears. That is because the smaller $\gamma(W)$, the harder the network is numerical controlled.

\begin{figure}
\centerline{\includegraphics[height=3in,width=5in]{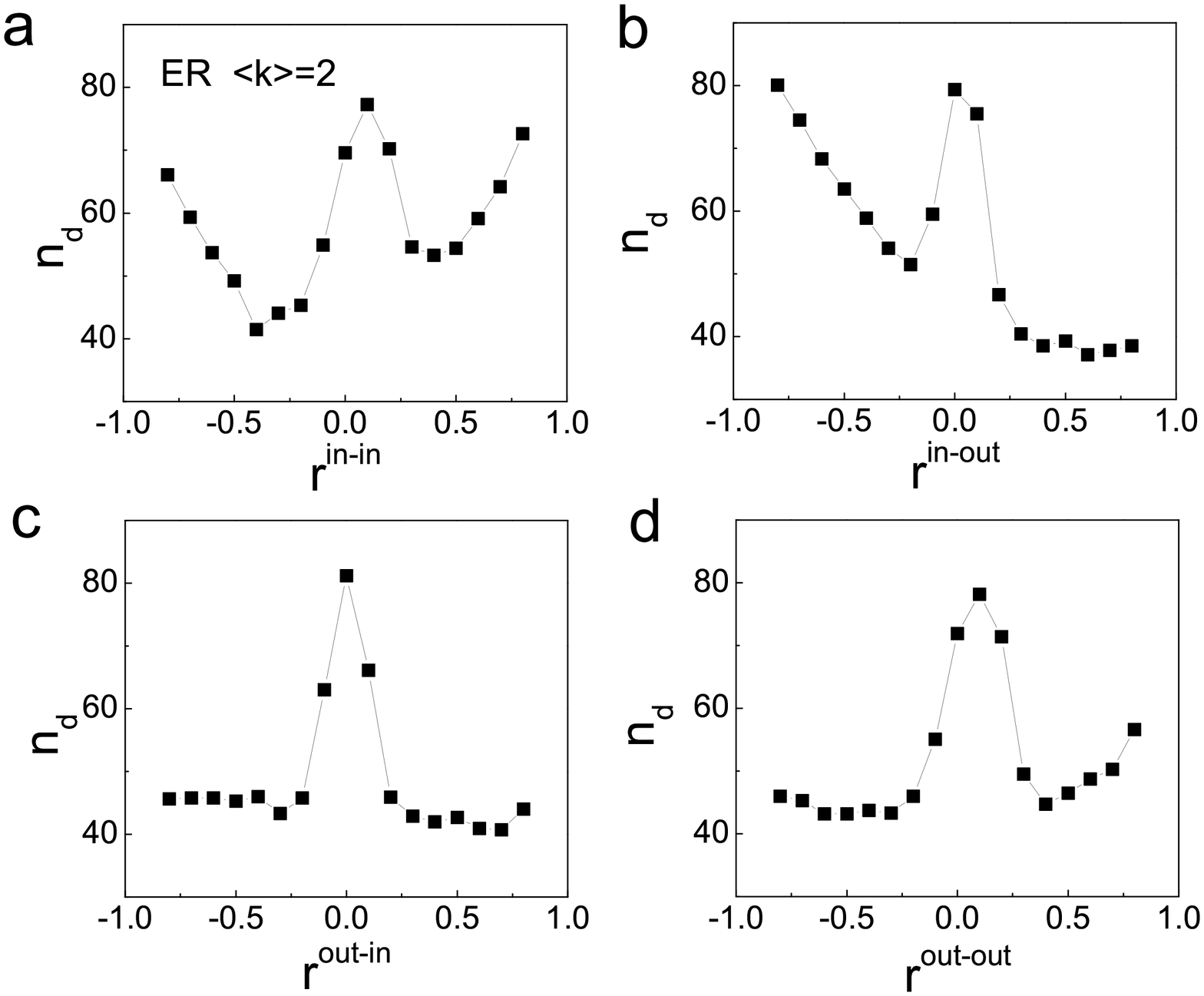}}
\caption{(Color online) The impact of degree correlations on the controllability transition points for ER networks with $N=100$ and average degree $<k>=2$.
} \label{fig4}
\end{figure}

\begin{figure}
\centerline{\includegraphics[height=3in,width=5in]{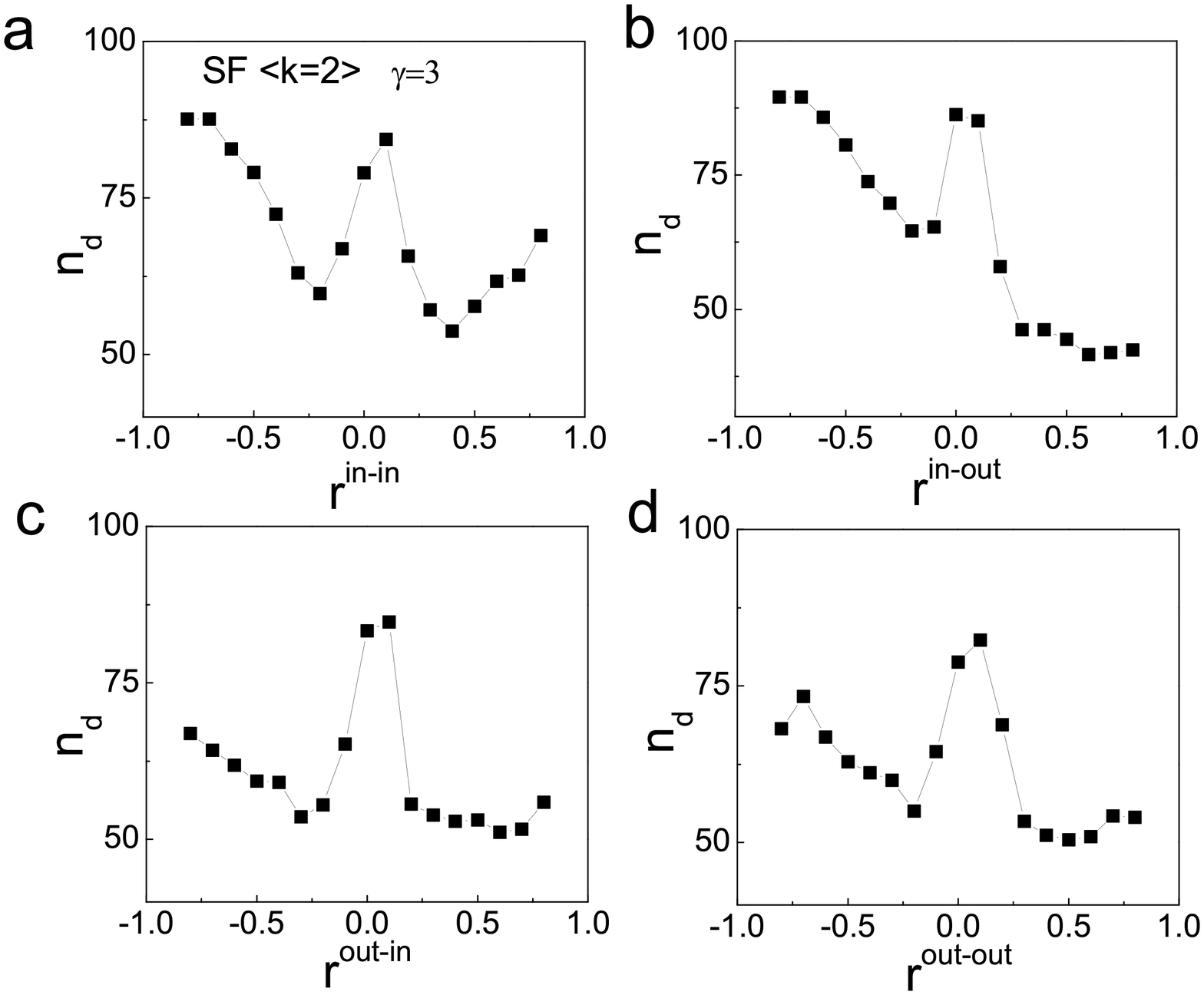}}
\caption{(Color online) The impact of degree correlations on the controllability transition points for SF networks with $N=100$, average degree $<k>=2$, and degree exponent $\gamma=3$.
} \label{fig5}
\end{figure}

 Next, it is necessary to explore the case of directed networks with self-loops. As shown in Fig.~\ref{fig4} and Fig.~\ref{fig5}, the transition points present the local maximum in both directed ER and SF networks as $r$ around $0$. The results are different from the situation in structural controllability\cite{posfai2013effect}, in which the number of driver nodes can produce linear, quadratic or no dependence on degree correlations.
  For ER networks, the value of transition points are larger as low (negative) correlations $r^{in-in}$ and $r^{in-out}$ while are smaller as both low (negative) and high (positive) correlations $r^{out-in}$ and $r^{out-out}$. For SF networks, the change of transition points shows the same tendency in Fig.~\ref{fig5}, which indicates the low (negative) correlations $r^{in-in}$ and $r^{in-out}$ impact the transition point greater.

\begin{figure}
\centerline{\includegraphics[height=2.8in,width=4in]{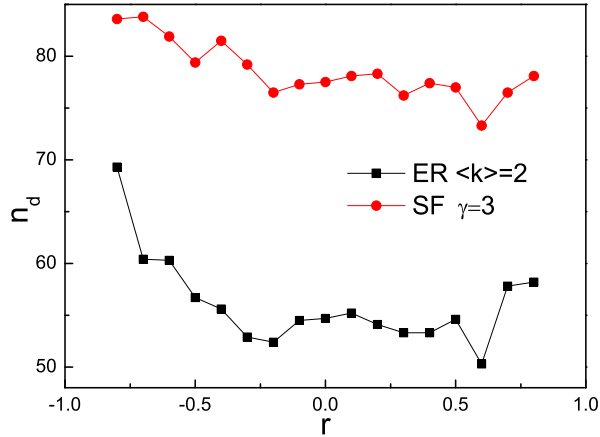}}
\caption{(Color online) The impact of degree correlations on the controllability transition points for undirected ER and SF networks without self-loops. The statistics and parameters not shown are the same used in Fig.1.
} \label{fig6}
\end{figure}

Finally, we examine how the self-loop affects controllability transition. Unlike taking no account of self-loop in structural controllability, the value of controllability transition point increases largely without self-loop in our simulation which are shown in Fig.~\ref{fig6}. The self-loop provides former information about state which makes the system easier controlled.

\section*{Discussion and Conclusion}

 The ultimate goal for us to study the complex networks is to control them.  Hence, a critical problem is whether the systems can be driven for an initial state to any desired state. A widely used formula is Kalman's controllability rank condition, however, Ref\cite{sun2013controllability} states that a system has a well conditioned Kalman's controllability matrix may
has an ill conditioned controllability Gramian in large networks, and the most important is the controllability success rate has a transition with the increasing of driver nodes. In addition, motivated by the fact that the degree correlation has important role in the dynamics on networks, we examine how the degree correlation affects the controllability transition points on networks.

 We use the numerical simulations to identify the variation of transition points in undirected and directed networks with varied degree correlations, finding the transition points appear local maximum around degree correlation $r=0.1$ and $r=0$ in sparse ER and SF networks respectively. The results also indicates that the transition point based on the minimal energy control cannot be influenced greatly by degree distribution but other parameters which relates to the control mode in dense networks.

In future research, we expect further studying the relationship between controllability and structure of complex networks and finding the proper criteria to determine the controllability in large systems.


\section*{Acknowledgments}

This work is funded by: The National Natural Science Foundation of China (Grant Nos.£º11275186, 91024026, FOM2014OF001).

\bibliography{ref}

\end{document}